\documentclass{ws-ijbc}

\usepackage{natbib}

\usepackage{multirow}
\usepackage{tabularx}
\usepackage{booktabs}
\usepackage{colortbl}
\usepackage{tikz}
\usepackage[toc,page]{appendix}

\usepackage{color}

\newcommand\ed{\stackrel{\op{d}}{=}}
\newcommand\R{{\mathbb R}}
\newcommand\E{{\mathbb E}}
\newcommand\PP{{\mathbb P}}
\newcommand\op[1]{\mathop{\rm #1}\nolimits}

\usepackage{graphicx}
\usepackage{dcolumn}
\usepackage{bm}
\usepackage{amsfonts}
\usepackage{amsmath, amssymb}

\title{Multifractal modeling of short-term interest rates}

\author{Martin Rypdal and Ola L\o vsletten \\
Department of Mathematics and Statistics, University of Troms\o, Norway \\ ~\\ November 2011
}

\begin{document}

\maketitle

\begin{abstract}
We propose a multifractal model for short-term interest rates. The model is a version of the Markov-Switching Multifractal (MSM), which incorporates the well-known level effect observed in interest rates. Unlike previously suggested models, the level-MSM model captures the power-law scaling of the structure functions and the slowly decaying dependency in the absolute value of returns. We apply the model to the Norwegian Interbank Offered Rate with three months maturity (NIBORM3) and the U.S. Treasury Bill with three months maturity (TBM3). The performance of the model is compared to level-GARCH models, level-EGARCH models and jump-diffusions. For the TBM3 data the multifractal out-performs all the alternatives considered. 
\end{abstract}

\section{Introduction} \label{Intro}

Interest rates play an important role for financial institutions, for instance in risk management. Interest-rate risk is often assessed via simple stress tests, where one considers parallel shifts in the yield curve (typically 100 or 200 basis points) and reports the changes in the value of the portfolio. This approach can be improved by stochastic modeling of future movements in the interest-rate yield curve. \citet{Vasicek1977} and 
\citet{Cox1985} argue that the yield curve is given by the spot rate alone. Short-term interest rates, such as NIBORM3 and TBM3, are frequently used as proxies for the spot rate, and hence accurate modeling of these time series is potentially very important. To our knowledge the present paper is the first published study considering the Norwegian rate, while the TBM3 has been
studied by e.g. \citet{Andersen1997}, \citet{Chapman2001}, \citet{Durham2003}, \citet{Johannes2004} and \citet{Bali2006}.

Traditionally, short-term rates have been modeled by It{\^o} stochastic differential equations on the form 
\begin{equation} \label{sde}
dR(s)=f(R(s) )ds+c R(s)^\gamma dB(s),
\end{equation}
where $s$ is the time variable, $R(s)$ denotes the risk-free rate and $B(s)$ is a Brownian motion. If we discretize this equation, by letting $s=t \Delta s$ and $r_t=R(s)$, then we obtain a stochastic difference equation on the form $r_{t}=r_{t-1}+f(r_{t-1})\Delta t+c \Delta s^{1/2}  r_{t-1}^\gamma  w_t$, where $w_t$ are independent and Gaussian distributed random variables with unit variance. It is convenient to write this equation on the form 
\begin{equation}
r_{t}=\mu_t+\sigma_t,
\end{equation}
with $\mu_t=r_{t-1}+\Delta s f(r_{t-1})$ and $\sigma_t=r_{t-1}^\gamma x_{t}$. Here $x_t=\sigma w_t$ with $\sigma=c \Delta s^{1/2}$. Throughout the paper we will consider linear drift terms on the form $f(r)=A_0+A_1r$. If $A_1<0$, then (\ref{sde}) has a stationary solution, and the discretized equation has a stationary solution for sufficiently small $\Delta s>0$. For a fixed $\Delta s>0$ we write
$\mu_t=r_{t-1}+\alpha_0+\alpha_1  r_{t-1}$. In this case we have stationarity for $-2<\alpha_1<0$.

\begin{figure}[h!]
\begin{center}
\includegraphics[width=14.0cm]{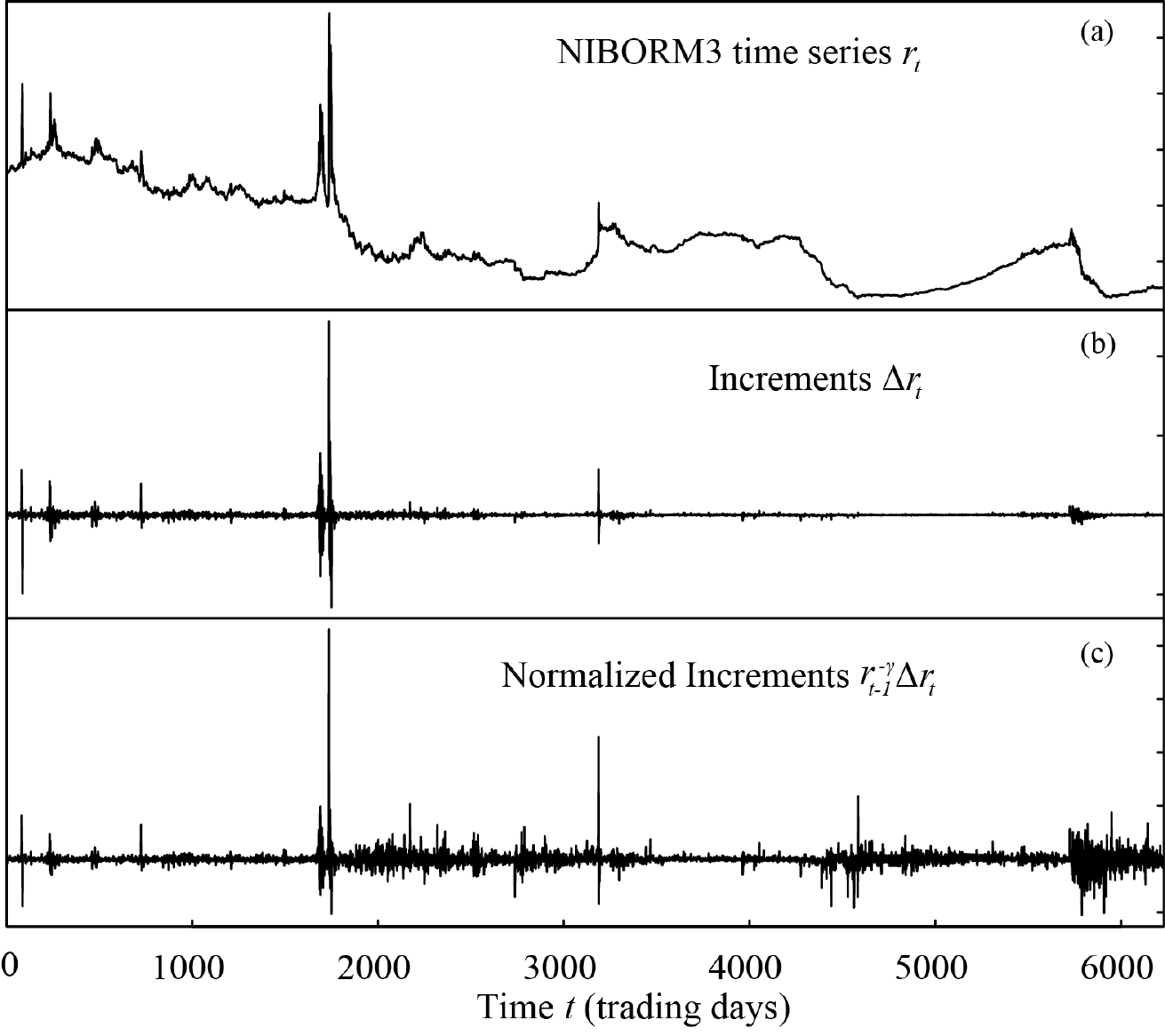}
\caption{(a): The NIBORM3 data $r_t$ for the time period 1986-01-02 to 2010-09-24. (b): The one-day increments $\Delta r_t=r_t-r_{t-1}$. (c): The normalized one-day increments $\Delta r_t/r_{t-1}^\gamma$. The value of the CEV parameter is $\gamma= 1.61$.  
} \label{fig1}
\end{center}
\end{figure}

\begin{figure}
\begin{center}
\includegraphics[width=14.0cm]{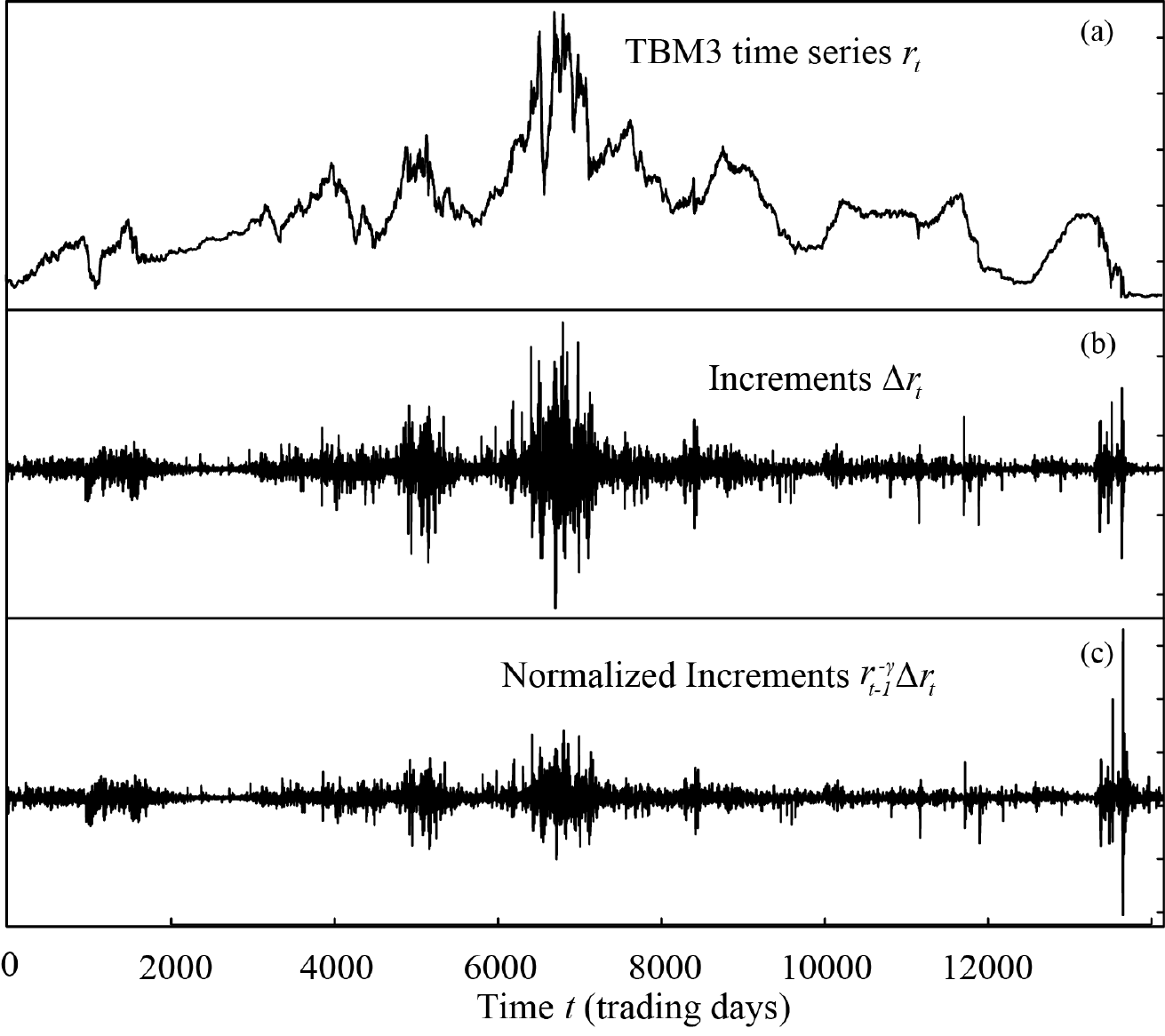}
\caption{(a): The TBM3 data $r_t$ for the time period 1954-01-04 to 2010-09-22. (b): The one-day increments $\Delta r_t=r_t-r_{t-1}$. (c): The normalized one-day increments $\Delta r_t/r_{t-1}^\gamma$. The value of the CEV parameter is $\gamma= 0.38$.  
} \label{fig2}
\end{center}
\end{figure}

The number $\gamma \geq 0$ is called the Constant Elasticity Variance (CEV) parameter. For $\gamma \neq 0$ the models feature the so-called level effect. It is generally accepted that this effect is present in interest-rate data, see for instance \cite{Longstaff1992}. The level effect introduces volatility persistence, i.e. strong dependence between the absolute values of the increments $\Delta r_t=r_{t}-r_{t-1}$. However, if $B(s)$ is a Brownian motion, then the variables
\begin{equation}
\frac{\Delta r_t-(\alpha_0+\alpha_1  r_{t-1} )}{r_{t-1}^\gamma}
\end{equation}
are i.i.d. and Gaussian, and hence the volatility persistence will vanish under a simple transformation of the increments. Such processes are called pure-level models and have been studied by \citet{Cox1985} and \citet{Longstaff1992}. Using time-series data for short-term interest rates we can optimize the likelihood to determine the parameters $\alpha_0$, $\alpha_1$ and $\gamma$. These results are shown in table \ref{tab2}. We can then transform the data according to (3). The resulting time series (which are plotted in figures \ref{fig1}(c) and \ref{fig2}(c)) are realizations of the so-called normalized increment process. Just from inspection of figures \ref{fig1}(c) and \ref{fig2}(c)) we can observe that the resulting time series exhibit strong volatility persistence. This is confirmed in figures \ref{fig1}(a) and \ref{fig1}(b), where we have plotted the autocorrelation functions for the absolute values of the normalized increments. As a consequence of these observations we conclude that pure-level models are insufficient, and we will therefore replace the process $w_t$ with dependent variables $x_t$. 
	
In our empirical investigations we find that the dependency structure in the variables $x_t$ resembles the stylized facts of logarithmic returns of asset prices, namely that the variables themselves are uncorrelated (or weakly correlated) whereas their absolute values have slowly decaying autocorrelation functions. To describe this dependency several authors (e.g. \citet{Brenner1996} and \citet{Koedijk1997})  have suggested using a generalized autoregressive conditional heteroskedasticity (GARCH) model for the process $x_t$. The corresponding process $r_t$ is then called a level-GARCH model. These models can be further improved by letting $x_t$ be an Exponential GARCH (EGARCH) model or a Markov Switching GARCH (MSGARCH) model. The level-EGARCH model was introduced by \citet{Andersen1997} and has been reported to perform better than the standard GARCH model on the TBM3 data. 

In this work we propose a multifractal model, specifically a level-MSM model, as an alternative to level-EGARCH models and level-MSGARCH models for short-term interest rates. This model is a slight modification of the standard MSM model constructed by Calvet and Fisher (2004). The motivation for introducing multifractal models for short-term interest rates is similar to the motivation for multifractal modeling of asset prices, namely that these models capture the dependency structure and scaling properties of the process $x_t$. Secondly, encouraging results (in- and out-of-sample) have been obtained in a preliminary study of the NIBORM3 data \cite{Loevsletten2010}. 
	
We follow Calvet and Fisher (2004) and use an adjusted version of the Vuong test \cite{Vuong1989} for model selection, and the main result of this work is that the level-MSM model performs well compared against level-EGARCH models, level-MSGARCH models and jump-diffusions. 
The paper is organized as follows: In section \ref{multifractals} we define the level-MSM model and consider some basic properties. Technical details about multifractal processes are presented in appendix \ref{examples}. A brief description of the alternative models are presented in section \ref{alternatives}, and in section \ref{results} the results of the Vuong test are presented. In section \ref{conclusion} we make some concluding remarks. \\

\begin{remark}{\em
The data analyzed in this paper is freely available online. Both data sets are given with daily resolution. The TBM3 data is taken from the period 1951-01-04 to 2010-09-22, and consists of 14172 data points. The NIBORM3 data is taken from the period 1986-01-02 to 2010-09-24, and consists of 6231 data points. The TBM3 data contains several days with zero value, which causes certain technical problems in the stochastic modeling. The problem is resolved by a simple variable shift $r \mapsto r+b$. In order to be consistent with our set of models, we estimate the constant $b>0$ by estimating the conditional standard deviations $\op{sdv}[\Delta r_t|r_{t-1}=r]$ (for various $r$) and fitting a function $c r^\gamma -b$ to this data set. With this approach we find $b=0.03$. 
}
\end{remark}

\section{Multifractal models} \label{multifractals}
The application of multifractal processes to finance was introduced by \citet{Mandelbrot1997}. A process $X(t)$ is multifractal if the structure functions 
$S_q (t)={\mathbb E}[|X(t) |^q]$ are power laws in $t$, and one define a scaling function $\zeta(q)$ through the relation $S_q (t) \sim t^{\zeta(q)}$ as $t \to 0$, i.e. 
$$
\zeta(q) = \lim_{t \to 0} \frac{\log S_q(t)}{\log t}\,.
$$ 
Using H{\"o}lder's inequality it is easy to show that the scaling function $\zeta(q)$ is concave, and we also note that if the process $X(t)$ is $h$-self similar, i.e. $X(at) \ed a^h X(t)$, then $\zeta(q)=hq$. We are therefore interested in the situations where $\zeta(q)$ is strictly concave. In this case we see that 
$$
\frac{\E[X(t)^4]}{\E[X(t)^2]^2} \sim t^{\zeta(4)-2 \zeta(2)} \to \infty \text{ ~~as~~ } t \to 0\,. 
$$
If we assume that $X(t)$ also has stationary increments, then this implies that the $\Delta t$-lagged increments $X(t+\Delta t)-X(t)$ are more leptokurtic for small $\Delta t$ than for larger $\Delta t$. In particular, $X(t)$ can not be a Gaussian process. 

In financial time series one will most often want the process $X(t)$ to have uncorrelated and stationary increments. 
In order for this to be satisfied we must impose the condition $\zeta(2)=1$, because otherwise the variables $\Delta X(t)=X(t+1)-X(1)$ have slowly decaying autocorrelation. In fact, for $\zeta(2) \neq 1$ 
$$
\op{Cov}(\Delta X(0) ,\Delta X(t)) \sim t^{2H_d-2}\,, 
$$
where $H_d:=\zeta(2)/2$. The advantage of a multifractal process (with strictly concave scaling function) is that it has strongly dependent increments, even for $H_d=1/2$. For instance, in the MSM model\footnote{It is pointed out by \citet{Lux2006} that the MSM model is only a finite-level approximation, and so equation (\ref{cov}) is only valid on time scales up to $b^K$. Here $b>0$ and $K \in {\mathbb N}$ are parameters in the MSM model. See section \ref{MSM}.} we have $\zeta(2)=1$ and 
\begin{equation} \label{cov}
\op{Cov}(|\Delta X(0)|^q, |\Delta X(t)|^q) \sim t^{-(\zeta(2q)-2 \zeta(q))}\,.
\end{equation}      
See proposition 1 in \cite{Calvet2004b}. This inherent long-range volatility persistence serves as our motivation for modeling short-term interest rates using multifractals.

\subsection{Mandelbrot's MMAR processes}
The multifractal models introduced by \citet{Mandelbrot1997} are stochastic processes on the form 
\begin{equation} \label{composition}
X(t)=\sigma\,T^{1/2}\, B \big{(} \Theta(t) \big{)}\,,
\end{equation}
defined for $ 0 \leq t \leq T$, where $\Theta(t)=\mu([0,t])$ is a random probability measure on $[0,T]$ and $B(t)$ is a Brownian motion with $\E[B(1)^2]=1$. 
The process $\Theta(t)$ is itself a multifractal process with scaling function $\zeta_\Theta(q)$, and if $\mu$ and $B(\cdot)$ are independent, then $\zeta_X(q)=\zeta_\Theta(q/2)$. By construction, the measures we will mention in this paper satisfy $\E[\mu([0,t])] \propto t$, i.e. $\zeta_\Theta(1)=1$. It follows that $\zeta_X(2)=1$. We note that processes with correlated increments are useful in other applications (see e.g. \cite{Rypdal2010}), and the Borownian motion $B(t)$ can in these situations be replaces by a fractional Brownian motion with $H \neq 1/2$.

There are several choices for the measure $\mu$, and for completeness we give three examples in the appendix. We also show how a particular construction, the Poisson multifractal, leads to a discrete-time MSM model defined in the next section.

\subsection{The MSM model} \label{MSM}

The Markov Switching Multifractal model of order $K>1$ is given by 
\begin{equation}
x_t=\sigma g(\mathbf{M}_t)^{1/2} \varepsilon_t \,,
\label{msm1}
\end{equation}
where ${\bf M}_t=(M_{1,t},\dots,M_{K,t})$, and the function $g:\mathbb{R}^K\rightarrow \mathbb{R}$ is the product of the vector components.
The innovatons $\varepsilon_t$ are Gaussian distributed with unit variance, and in this setting also assumed to be independent.  
The process $\mathbf{M}_t$ is a Markov chain defined by the following updating scheme:
\begin{equation}
M_{k,t} = \begin{cases}
\text{drawn from }M &\text{with probability } \lambda_k,\\
M_{k,t-1} &\text{with probability } 1-\lambda_k.
\end{cases}
\label{msm2}
 \end{equation}
Components are updated independently of all previous updates, and the frequencies $\lambda_k$ are related to each other through (\ref{lambda}). 

For simplicity we choose a two-point distribution $M$:
\begin{equation}
 \mathbb{P}(M=m_0)=\mathbb{P}(M=m_1)=\frac{1}{2}\,, \text{ with ~} m_0\in(1,2) \text{~ and ~} m_1=2-m_0.
\end{equation}
This version of the MSM model is known as the binomial multifractal \cite{Calvet2004b}. To higlight the number of multipliers $K$ we will use the notation MSM($K$). Note that, unlike GARCH$(p,q)$, the number of parameters remains unchanged with increasing order.   

Let $\{{\bf m}^i\,|\,i=1,\dots,d:=2^K\}$ denote the sample space of ${\bf M}_t$. We also define vectors $\mathbf{\omega}(x):\mathbb{R}\rightarrow \mathbb{R}^{d}$ and $\mathbf{p}_t:\mathbb{N}\rightarrow \mathbb{R}^{d}$ with components $\omega_k(x)=n(x|\sigma^2g(\mathbf{m}^k))$ and $p_{k,t}=\mathbb{P}(\mathbf{M}_{t}=\mathbf{m}^k|\mathbf{x}_{t})$ respectivly. Here ${\bf x}_t = (x_2,\dots,x_t)$, and $n(\cdot|\sigma)$ is the density of the normal distribution with zero mean and variance $\sigma^2$. With this notation the density of $x_t|\mathbf{x}_{t-1}$ is given by 
\begin{equation} \label{L1}
p(x_t|\mathbf{x}_{t-1})=\mathbf{p}_{t-1}^TA\,\mathbf{\omega}(x_t),
\end{equation}
where $A$ is the transitition matrix of the Markov chain, that is $A_{j,k}=\mathbb{P}(\mathbf{M}_t=\mathbf{m}^k|\mathbf{M}_{t-1}=\mathbf{m}^j)$. This is seen by conditioning on the underlying Markov-chain at time $t-1$ and $t$, and using Bayes' rule. It is easily seen, once again using Bayes' rule, that the vectors $\mathbf{p}_t$ follow the recursion
\begin{equation} \label{L2}
\mathbf{p}^T_t=\frac{1}{p(x_t|\mathbf{x}_{t-1})}\mathbf{\omega}^T(x_t)*
(\mathbf{p}^T_{t-1}A),
\end{equation}
where $*$ denotes the Hadamard product. We start the recursion with the limiting probabilities of $\mathbf{M}_t$. 

\subsection{The level-MSM model for interest rates}
We propose the following level-MSM model for short-term interest rates: 
\begin{equation} \label{levelMSM}
r_t-r_{t-1}=\alpha_0+\alpha_1 r_{t-1} + r_{t-1}^\gamma x_t,
\end{equation}
where $x_t$ is the binomial MSM model defined above. 
The likelihood 
$$
L=\sum_{t=2}^n \log f(r_t|{\bf r}_{t-1})\,,
$$
 for data $r_2,\ldots,r_n$, with $r_1$ taken as a pre-sample value, now follows from equations (\ref{L1}) and (\ref{L2}) together with the relation 
$$
f(r_t|{\bf r}_{t-1}) = \frac{1}{r_{t-1}^\gamma} \, p(x_t|\mathbf{x}_{t-1})\,,
$$ 
where $f$ is the density of $r_t|{\bf r}_{t-1}$. 

We have fitted the level-MSM models to the two time series under consideration for $K=2,\dots,9$. 
For the case $K=9$, the maximum likelihood (ML) estimates for the parameters $\alpha_0$ and $\alpha_1$ are reported in table \ref{tab8}. In this table the estimates of $\alpha_0$ and $\alpha_1$ for the alternative models are also included. As expected the estimates for the parameter $\alpha_1$ are negative, but we observe that none of these ML estimates  are significantly different from zero for the TBM3 data.  The estimates for $\alpha_1$ are also very small for the NIBORM3 data, and we will therefore consider the model defined by (\ref{levelMSM}) with $\alpha_1=0$. The corresponding ML estimates for the other parameters are presented in table \ref{tab3}.  From these results we also see that the likelihoods increase monotonically with $K$. Using a Vuong test we compare the level-MSM of order $K=9$ against level-MSM models of lower order. When comparing two models with this method, the null-hypothesis is that both models are equally far from the data-generating process measured by the Kullback-Leibler distance. Hence small $p$-values indicate that the level-MSM model of order $K=9$ is significantly closer to the true generating process than the level-MSM models of order $K=7$ or lower. 

The same model-selection-test is used to compare the level-MSM model with various alternative models, and more details on the test are presented in section \ref{results}.

\color{black}

\section{Alternative models} \label{alternatives}
In this section we briefly discuss some processes which we use for benchmarking the level-MSM model. As for the level-MSM model we have first considered these models with drift terms on the form $\mu_t=r_{t-1}+\alpha_0+\alpha_1 r_{t-1}$. The ML estimates for the parameters $\alpha_0$ and $\alpha_1$ are reported in table \ref{tab8}. Again we see that the contribution from the parameter $\alpha_1$ is negligible. Consequently we will consider models with drift terms on the form $\mu_t=r_{t-1}+\alpha_0$.

\subsection{The level-GARCH model}
In this paper we prefer the standard level-GARCH model proposed by \citet{Koedijk1997}. As is common for this model, we include student-$t$ distributed innovations. The model then reads  
\begin{equation} \label{levelGARCH}
r_t=\alpha_0 +r_{t-1}  +\sigma_t, \quad
\sigma_t=h_t^{1/2} r_{t-1}^\gamma\varepsilon_t,\quad  h_t=a_0+a_1x_{t-1}^2 +b\,h_{t-1},
\end{equation}
where the innovations $\varepsilon_t$ are i.i.d. $ t_{\nu}(0,1)$\footnote{$ t_{\nu}(0,1)$ is a centralized, unit-variance student-$t$ distribution with $\nu$ degrees of freedom.}. For $\gamma=0$ we have the pure-GARCH(1,1) model, and for $a_1=b=0$ we have the standard level model.

The estimated parameters for the NIBORM3 and TBM3 data are presented in table \ref{tab4}. We observe that simulated paths with these exponents have far too wild fluctuations compared to the real data, indicating that this model fails to accurately describe the interest-rate fluctuations. In addition, we know that the GARCH models exhibit exponentially decaying autocorrelation functions for the absolute values of the increments. This means that the long-range volatility persistence observed in the short-term interest rates is not inherent in these models.    

\subsection{The level-EGARCH model}

Instead of using (\ref{levelGARCH}), \citet{Andersen1997} propose using the EGARCH model. They find that this model gives an adequate fit to the TBM3 data, and a better fit compared to the standard GARCH model. 
In the EGARCH model the logarithm of the conditional variance replaces the conditional variance. The variance-recursion is then
\begin{equation}
\log h_t = a_0 +a_1\varepsilon_{t-1}+a_2|\varepsilon_{t-1}|+b\log h_{t-1}.
\end{equation}
The extra parameter $a_1$ controls potentially different responses to positive and negative returns. The use of the logarithm guarantees positive values of the volatility for all parameter values. In addition, this model provides the extra flexibility by letting the conditional distribution be non-symmetric. As we will become apparent from the results presented in the next section, the EGARCH model gives better results than the GARCH model for both of the time series considered in this paper. 
This confirms the results of \citet{Andersen1997}. 

The ML estimates for the parameters in the EGARCH model are presented in table \ref{tab5}.



\subsection{Jump-diffusions}
The final class of alternative models considered are the jump-diffusions. Both \citet{Johannes2004} and \citet{Das2002} propose these processes in order to describe the large spikes observed in interest-rate data. As a discretized jump-diffusion model we use the following specification:
\begin{equation*} 
r_t=\alpha_0+r_{t-1}+ r_{t-1}^\gamma (x_t + J_t z_t), 
\end{equation*}
where $J_t$ is the jump-indicator assumed to follow a Bernoulli-distribution. The probability of a jump taking place at time $t$ is then given by $\left(1+\exp(-c-d r_{t-1})\right)^{-1}$. By letting $x_t=h_t^{1/2}\varepsilon_t$ follow a GARCH-process, conditional heteroschedacity in levels is also accomodated \cite{Das2002,Hong2004}. 
The innovations are distributed as $z_t \stackrel{\op{d}}{\sim} {\mathcal N}(0,\tau^2)$ and $\varepsilon_t \stackrel{\op{d}}{\sim} {\mathcal N}(0,1)$.The variance recursion in the GARCH process is given by
\begin{equation*} 
h_t =a_0 + a_1 \left(\frac{\Delta r_{t-1}-\alpha_0}{r_{t-2}^\gamma }\right)^2+ b\, h_{t-1},
\end{equation*}
where $\Delta r_{t}=r_t-r_{t-1}$. 

The ML estimates for the parameters in the jump-diffusion model are presented in table \ref{tab6}.

\color{black}

\section{In-sample comparision} \label{results}

To test the binomial MSM model against the alternative models we emply a version of the Vuong test which is adjusted for heteroschedacity and autocorrelation (HAC)  \cite{Calvet2004b}.    
For each of the alternative models the null-hypothesis is that this model and the level-MSM model of order $K=9$ are equidistant from the true data-generating process, measured in the  
Kullback-Leibler distance \cite{Kullback1951}. In the classical Vuong test it is assumed that the data-generating process is i.i.d., and then the log-ratio of the likelihoods for the two models will converge to a normal distribution with zero mean, and with a variance which is consistently estimated by the sample variance for the log-ratio of the likelihoods.  
Using the corresponding normal distribution one can then easily calculate a $p$-value under the null-hypothesis. In the HAC-adjusted version of this test, the data-generating process may exhibit dependence, but the variables $r_t$ should be identically distributed. This is satisfied if we assume that the data-generating process $r_t$ is stationary.  

The results of this test are presented in table \ref{tab7}. We observe that for the TBM3 data, the level-MSM model of order $K=9$ is significantly closer to the data-generating process than any of the alternative models. For the NIBORM3 data the EGARCH model performs best, whereas the level-MSM model performs better than the standard level-GARCH model and jump-diffusions.

\section{Concluding remarks} \label{conclusion}
In this paper we have introduced a multifractal model for short-term interest rates. The model combines the well-established level effect described in \cite{Longstaff1992} with the discretized multifractal model of \cite{Calvet2001}. In a comparison with level-GARCH, level-EGARCH and jump-diffusions, we find that this model well describes the fluctuations of the TBM3 and NIBORM3 time series. The main result of this work is that the level-MSM outperforms all alternatives for the TBM3 data. This motivates further research on multifractal modeling of short-term interest rates, in particular an out-of-sample analysis of the level-MSM model.  

It is also interesting to note in the level-GARCH model the
parameter estimates (for both the TBM3 data and the NIBORM3 data) fall
outside the covariance-stationarity region. 
As a result the level-GARCH model has a wild volatility pattern, which does not seem to be an accurate description of the interest-rate data. This confirms the results of e.g. \citet{Andersen1997}.     
 ~\\
 
 \begin{appendix}{Examples of multifractal measures}\label{examples}

In this appendix we give examples of random measures that can be used to define multifractal processes via (\ref{composition}). For the Poisson multifractal we show how a discretization leads to the model in (\ref{msm1}). \\

\noindent {\bf Example 1:}
In the simplest case $\mu$ is a randomized dyadic Bernoulli measure with probabilities $p_1=p$ and $p_2=1-p$. This measure is constructed through an iterative procedure, where we in the first step divide the interval $[0,T]$ in two pieces $\Delta_0$ and $\Delta_1$ of equal length. One of the intervals is chosen at random and given a mass $p_1 \,\mu([0,T])$, while the other interval is given the mass $p_2 \,\mu([0,T])$. This procedure is then repeated recursively,  i.e. the interval $\Delta_0$ is divided into the equally sized intervals $\Delta_{00}$ and $\Delta_{01}$. One of these intervals is given the mass $p_1^2 \,\mu([0,T])$ and the other is given the mass $p_1 p_2 \,\mu([0,T])$. Formally the measure can be defined by letting $f$ be a random bijection\footnote{With probability $1/2$ for each of the two outcomes $(0,1) \mapsto (p_1,p_2)$ and $(0,1) \mapsto (p_2,p_1)$} $\{0,1\} \to \{p_1,p_2\}$ and $f_{i_1,\dots,i_k}$ be independent copies of $f$. The measure is constructed by assigning the mass $\mu(\Delta_{i_1,i_2,\dots,i_k})=f(i_1)f_{i_1}(i_2) \cdots f_{i_1,i_2,\dots,i_{k-1}}(i_k)$ 
to the dyadic intervals 
$$
\Delta_{i_1,\dots,i_k}=\big{[}T \cdot (0.i_1 \cdots i_k)_2, T \cdot (0.i_1 \cdots i_k)_2+T\cdot 2^{-k}\big{]}\,,\,\,\,\,i_n \in \{0,1\}\,.
$$
A simple combinatory argument shows that for $t=T \cdot 2^{-k}$, the random variable $\Theta(t)$ has density 
$$
p_{\Theta,t}(x)=\frac{1}{2^k} \sum_{n=0}^k \binom{k}{n} \delta(x-p^n(1-p)^{k-n})\,.
$$
When the processes $B(t)$ and $\Theta(t)$ are independent, we obtain the density of $X(t)$: 
$$
p_{X,t}(x)=\int p_{\sigma\,T^{1/2}\,B,s}(x)\,p_{\Theta,t}(s)\,ds = \frac{1}{2^k} \sum_{n=0}^k \binom{k}{n} 
\frac{\exp \Bigg{(} -\frac{x^2}{2 \sigma^2 \big{(}Tp^n(1-p)^{k-n}\big{)}}\Bigg{)}}{\sqrt{2 \pi \sigma^2 \big{(}Tp^n(1-p)^{k-n}\big{)}}} \,.
$$
From this density we can easily calculate the structure functions $S_q(t)=\E[|X(t)|^q]$ and see that $S_q(t) \sim t^{\zeta(q)}$ as $t \to 0$, where 
$\zeta(q)=1-T(q/2)$ and $T(q)=\log_2 ( p^q+(1-p)^q )$. 

\begin{remark} 
{\em The function $T(q)$ is sometimes called the scaling function of the random measure. The dyadic measure defined by $\mu(\Delta_{i_1,\dots,i_k})=p_{i_1} \cdots p_{i_k}$ has a multifractal spectrum $f(\alpha)$ given by the Legendre transform of $T(q)$: $f(\alpha)=\inf_{q \in \R} \{\alpha q-T(q)\}$, and the Hentchel-Procaccia dimension spectrum is $D_q=(q-1)^{-1} T(q)$. See e.g. \cite{Pesin1997}  for an account of the relation between scaling functions and spectra of fractal dimensions. The function $\alpha \mapsto f(\alpha/2)$ is often called the singularity spectrum of the process $X(t)$.} 
\end{remark}

\noindent {\bf Example 2:} A different class of multifractal measures are the $b$-adic random multiplicative cascades (see e.g. \cite{Mandelbrot1997} for a more detailed account). For an integer $b \geq 2$ and $i_n \in \{0,1,\dots,b-1\}$ we define the $b$-adic subintervals of $[0,T]$ as 
$$
\Delta_{i_1,\dots,i_k} = \big{[}T \cdot (0.i_1 \dots i_k)_b\,,\,T \cdot (0.i_1 \dots i_k)_b+T \cdot b^{-k}\big{]}\,.
$$
For a positive random variable $M$, with $\E[M]=b^{-1}$, let
$$
\mu(\Delta_{i_1,\dots,i_k}) = M_{i_1}M_{i_1i_2} \cdots M_{i_1i_2 \cdots i_k}\,\Omega_{i_1,\dots,i_k}\,,
$$ 
where $M_{i_1,\dots,i_n}$ are independent copies of $M$ and 
$$
\Omega_{i_1,\dots, i_k} = \lim_{n \to \infty} \sum_{j_1,\dots,j_n} M_{i_1} \cdots M_{i_1i_2 \cdots i_k} M_{i_1i_2 \cdots i_k j_1} \cdots M_{i_1i_2 \cdots i_k j_k \cdots j_n}\,.
$$

Again a stochastic process is constructed according to equation (\ref{composition}) with $\Theta(t)=\mu([0,t])$. This process has a scaling function $\zeta(q)=1-T(q)$, where 
$T(q)=\log_b \E[M^q]$. Popular choices for the multiplier $M$ are log-normal distributions, which give quadratic scaling functions, or other log-infinitely divisible distributions.
Almost every realization of the measure $\mu$ has the multifracal spectrum $f(\alpha)=\inf_{q \in \R} \{\alpha q-T(q)\}$. \\

\noindent {\bf Example 3:}
The Poisson multifractal measure on $[0,T]$ generalizes the $b$-adic multiplicative cascade by introducing randomness in the construction of the intervals $\Delta_{i_1, i_2, \dots , i_k}$. In the original multifractal models, each interval on level $k$ are divided into $b$ pieces of equal length at level $k+1$. As a result the interval length decreases as $b^{-k}$ with the level $k$. In the Poisson multifractal measure, the splitting of an interval at level $k$ is preformed by drawing the lengths of the new pieces randomly from an exponential distribution with rate $l_{k+1}$. This means that the mean and median length of an interval at level $k+1$ are $1/l_{k+1}$ and $\log(2)/l_{k+1}$ respectively, so to maintain exponential decay of interval lengths (as a function of level), one chooses $l_{k}=b^{k-1} l_1$. Note that $b$ no longer is restricted to the integers.

Formally, the Poisson multifractal measure on the interval $[0,T]$ is defined via a sequence of measures $\mu_k$ specified on randomly generated intervals 
$$
\Delta_{i_1, i_2, \dots , i_k} =[t_{i_1, i_2,  \dots , i_k}, t_{i_1, i_2,  \dots , (i_{k}+1) }]\,,
$$
where the numbers $t_{i_1, \dots, i_k}$ are defined by the following recursive construction:
Let $\{ \tau_{i_1,\dots,i_k}(j) \}_{j \in {\mathbb N}}$ denote independent and exponentially distributed random variables with rates $l_{k}=b^{k-1} l_1$. Define 
\begin{equation} \label{N}
N_{i_1,\dots,i_{k-1}}=\max \Big{\{} m : \sum_{j=1}^{m} \tau_{i_1,\dots,i_{k-1}}(j) < \op{diam} (\Delta_{i_1, i_2, \dots , i_{k-1}}) \Big{\}}
\end{equation}
and
\begin{equation} \label{t}
t_{i_1,\dots,i_k} = 
\begin{cases}
t_{i_1,\dots,i_{k-1}} & \text{ if }\,\,\, i_n=0 \\ 
t_{i_1,\dots,i_{k-1}} + \sum_{j=1}^{i_k} \tau_{i_1,\dots,i_{k-1}}(i_k) & \text{ if }\,\,\, 1 \leq i_k \leq N_{i_1,\dots,i_{k-1}} \\ 
t_{i_1,\dots,(i_{k-1}+1)} & \text{ if }\,\,\, i_k=N_{i_1,\dots,i_{k-1}} + 1
\end{cases}\,.
\end{equation}
This means that the interval $\Delta_{i_1,\dots, i_k}$ is divided into $N_{i_1,\dots,i_k}$ subintervals by the cuts made by a Poisson process with rate $l_{k+1}$. We start with an interval $\Delta=[0,T]$. 
A sequence $\mu_k$ of measures can now be defined via the formula 
$$
\mu_k(\Delta_{i_1, i_2, \dots , i_k}) = T^{-1}\,\op{diam}(\Delta_{i_1, i_2, \dots , i_k})\,M_{i_1} M_{i_1,i_2} \cdots M_{i_1, \dots, i_k}\,,
$$
where $M_{i_1 \cdots i_k}$ are independent copies of a positive random variable $M$ satisfying $\E[M]=1$. If $\E[M^2]<b$, then $\mu_k$ converges weakly to a Borel measure $\mu$ on $[0,T]$. 
With this choice of random measure the model given by equation (\ref{composition}) has scaling function 
\begin{equation} \label{Pscaling}
\zeta(q)=1-T(q/2)\,, \text{~~ where ~~} 
T(q)=1-q+\log_b \E[M^q]\,.
\end{equation}

Calvet and Fisher constructed their MSM model by discretizing the time interval $[0,T]$ and assigning discrete geometric (rather than an exponential) distributions on the waiting times $\tau$. More precisely one will fix an integer $K>1$ determining the number of levels that are to be included in the discrete model, and consider the integer values $\{0,1,\dots,\tilde{T}\}$, where $\tilde{T}=m^K$ for some positive integer $m$. As for the construction of the Poisson multifractal one makes random partitions $\Delta_{i_1, i_2, \dots , i_k} =[t_{i_1, i_2,  \dots , i_k }, t_{i_1, i_2,  \dots , (i_{k}+1) }]$ of the interval $[0,\tilde{T}]$. These are constructed using the recursive procedure described by equations (\ref{N}) and (\ref{t}), where the variables $\tau_{i_1,\dots,i_k}$ are replaced by discrete random variables $\tilde{\tau}_{i_1,\dots,i_k}$ with geometric distributions   
$\PP(\tau_{i_1,\dots,i_k} = \tau')=\lambda_k\,(1-\lambda_k)^{\tau'-1}$, 
with $1-\lambda_k=e^{- l  b^{k-1} \tilde{T}/T }$. This choice of parameters $\lambda_k$ implies that the median of $\tilde{\tau}_{i_1,\dots,i_k}$ is proportional to $l^{-1} b^{-n}$. 
Note that the parameters $\lambda_k$ are given by $\lambda_1$ through the formula 
\begin{equation} \label{lambda}
\lambda_k=1-(1-\lambda_1)^{b^{n-1}}\,.
\end{equation} 

A measure $\tilde{\mu}$ is defined by specifying that if $t$ is an integer and $[t-1,t] \subseteq \Delta_{i_1,\dots,i_{K}}$, then 
$$
\tilde{\mu} ([t-1,t]) = m^{-K} M_{i_1}M_{i_1,i_2} \cdots M_{i_1,\dots, i_K}\,,
$$
where $M_{i_1,\dots,i_k}$ are independent copies of a positive random variable $M$ with $\E[M]=1$. Again we define
$\tilde{\Theta}(t)=\tilde{\mu}([0,t])$, and by taking the composition with a fractional Brownian motion we get a discrete-time stochastic process
$\tilde{X}(t)=C \,B(\tilde{\Theta}(t))$. We denote $x_t=\tilde{X}(t)-\tilde{X}(t-1)$, and observe that 
\begin{eqnarray*}
x_t \ed C  \,\tilde{\mu}([t,t-1])^{1/2}\,\Big{(}B(t)-B(t-1) \Big{)} = \sigma\, (M_{i_1}M_{i_1,i_2} \cdots M_{i_1,\dots, i_K})^{1/2}\,\varepsilon_t\,,
\end{eqnarray*}
where $\sigma=C\,m^{-K}$ and $\varepsilon_t=B(t)-B(t-1)$ is a discrete version of a white Gaussian noise. We can simplify notations by denoting $M_{k,t}=M_{i_1,\dots,i_k}$ for $t \in \Delta_{i_1,\dots,i_k}$.  

We finally remark that there exists a large class of multifractal random measures, knowns as log-infinitely divisible cascades \cite{Bacry2001,Bacry2003,Bacry2008}, which have multifractal scaling and stationary increments. However, there only exists approximate maximum likelihood methods for these processes \cite{Lovsletten2011}, and hence the Voung-testing preformed in this paper is not available.   

~ \\
\end{appendix}

\noindent {\bf Acknowledgment.}
This project was partly funded by {\em Sparebank 1 Nord-Norge} and the Norwegian Research Council (project number 208125).

\newpage

\section{Figures and tables}

\noindent
\newpage
\noindent
\newpage
\noindent
\begin{figure}
\begin{center}
\includegraphics[width=14.0cm]{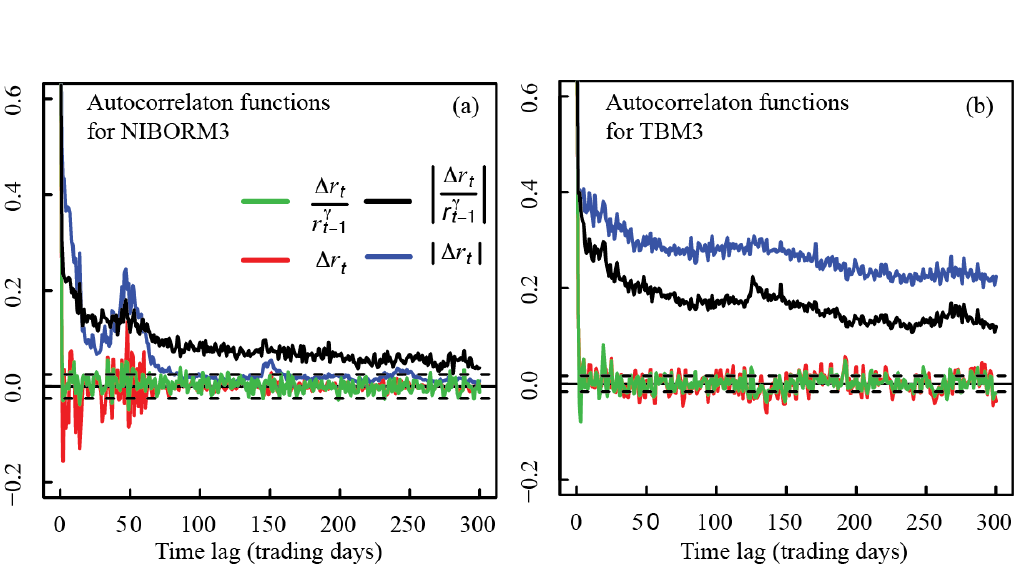}
\caption{(a): Autocorrelation functions for one-day increments in the NIBORM3 data. The variables considered are the standard increments $\Delta r_t=r_{t}-r_{t-1}$, the normalized increments $\Delta r_t/r_{t-1}^\gamma$ and the absolute values of these two time series. The value of the CEV parameter is $\gamma= 1.61$. (b): Same as (a), but now for the TBM3 data. The value of the CEV parameter is $\gamma= 0.38$. In both figure the dotted lines represent a 0.95 confidence interval for the autocorrelation function assuming independence. 
} \label{fig3}
\end{center}
\end{figure}

\newpage


\begin{table}
\begin{center}
\begin{minipage}{70mm}
\tbl{CEV-models.}
{\begin{tabular}{l  c c  c c c}
\toprule
Innovations &  $\alpha_0$ & $\gamma$ &$\sigma$  &$\nu$ &  $\log L$   \\	
\colrule 
\multicolumn{6}{c}{ U.S. Treasury Bill}\\
Normal&  $ 0.0009709 $ & $0.3755  $ &$0.04788$  & &  $ 15707.94$   \\	
$t_\nu$ &  $-0.0002434  $ & $0.5377  $ &$0.2186 $  &$2.01$ &  $20688.54$   \\	
\multicolumn{6}{c}{ NIBORM3}\\
Normal&  $0.0002863 $ & $1.61 $ &$0.006298$  &$$ &  $4771.78$   \\	
$t_\nu$ &$0.0002419 $ &  $0.988 $ & $0.0921 $ &$2.01$   &  $8041.42$   \\	
\botrule
\end{tabular}}
\tabnote{The constant elasticity volatility (CEV) model of \cite{Longstaff1992} with normal and student-$t$ innovations.} \label{tab2}
\end{minipage}
\end{center}
\label{symbols}
\end{table}

\begin{table}
\begin{center}
\begin{minipage}{\linewidth}
\tbl{Multifractal models}
{\begin{tabular}{r c  c c c c c c c c c}
\toprule
$K$& $10^3 \times\alpha_0 $ &$\gamma$ & $m_0$ & $b$  
& $\lambda_K$  & $\sigma$ & $\log L$ & BIC & Vuong & HAC-adj.\\
\colrule
\multicolumn{11}{c}{ U.S. Treasury Bill M3 } \\
$2$& $0.2752  $ &$0.2824 $ & $1.808 $ & $23.85  $  
& $0.1391 $  & $0.07037
$ & $22309.49$ & -3.1446 &   13.573& 9.091\\
$$& $(0.247 ) $ &$(0.0106 )$ & $(0.00471 )$ & $(4.92   )$  
& $(0.011 )$  & $(0.00162 )$ & $$ &  &  ($<0.001$)& ($<0.001$)\\

$3$& $ 0.2052 $ &$0.1913 $ & $1.73 $ & $10.62  $  
& $0.1660 $  & $0.08078$ & $22676.53
$ & -3.1964 & 11.618& 10.138\\
$$& $(0.234) $ &$(0.0175 )$ & $(0.0062 )$ & $(1.45  )$  
& $(0.0163 )$  & $(0.00336)$ & $$ &  &($<0.001$)  & ($<0.001$)\\

$4$& $0.2276  $ &$0.2269 $ & $1.661 $ & $9.643 $  
& $0.2956 $  & $0.07202$ & $22818.9 
$ & -3.2165 & 8.4076 &  7.562\\
$$& $(0.226 ) $ &$(0.032 )$ & $(0.00674 )$ & $(1.23  )$  
& $(0.0462 )$  & $(0.00357)$ & $$ &  &  ($<0.001$)& ($<0.001$)\\

$5$& $0.0643  $ &$0.2280 $ & $1.609 $ & $9.804  $  
& $0.6761 $  & $0.06865$ & $22901.03 $ & -3.2281 &5.036
 &  4.805\\
$$& $(0.222 ) $ &$(0.0191)$ & $(0.00667 )$ & $(1.04 )$  
& $(0.0597 )$  & $(0.00265)$ & $$ &  & ($<0.001$) & ($<0.001$)\\

$6$& $0.0375 $ &$0.2115 $ & $1.559 $ & $6.577   $  
& $0.778 $  & $0.06368
$ & $22920.57 $ & -3.2308 &  4.346 & 4.335\\
$$& $(0.219 ) $ &$(0.0282 )$ & $(0.00744 )$ & $(0.564  )$  
& $(0.0541 )$  & $(0.00360)$ & $$ &  & ($<0.001$) & ($<0.001$)\\

$7$& $0.0660 $ &$0.1744 $ & $1.527 $ & $5.713  $  
& $0.7928 $  & $0.08208$ & $22933.55 $ & -3.2326 & 3.4883
 & 3.112 \\
$$& $(0.223 ) $ &$(0.0291 )$ & $(0.00752 )$ & $(0.465  )$  
& $(0.0561 )$  & $(0.00441)$ & $$ &  & ($<0.001$) & (0.001)\\

$8$& $0.0839  $ &$0.1575 $ & $1.503 $ & $4.932 $  
& $0.8474 $  & $0.06956$ & $22951.89 $ & -3.2352 & 1.131 &0.938 \\
$$& $(0.215) $ &$(0.0260  )$ & $(0.00815  )$ & $(0.355 )$  
& $(0.0509 )$  & $(0.0036)$ & $$ &  &  (0.129)& (0.174)
\\

$9$& $0.0693  $ &$0.1984 $ & $1.462 $ & $3.864   $  
& $0.931 $  & $0.06029$ & $22955.80$ & -3.2358 &  &\\
$$& $(0.215 ) $ &$(0.037 )$ & $(0.00724 )$ & $(0.255  )$  
& $(0.0396 )$  & $(0.00443)$ & $$ &  &  & \\
\colrule
\multicolumn{11}{c}{ NIBORM3} \\

$2$& $1.156  $ &$1.805 $ & $1.843 $ & $16.23  $  
& $ 0.1016 $  & $0.006092$ & $8514.45$ & -2.7250
 & 5.219 & 5.298\\
$$& $(0.408  ) $ &$(0.04 )$ & $(0.00547 )$ & $(4.68 )$  
& $(0.0116 )$  & $(0.000516)$ & $$ &  & ($<0.001$) &($<0.001$)\\

$3$& $1.041  $ &$1.761 $ & $1.772 $ & $18.37  $  
& $0.4259 $  & $0.005674$ & $8671.87$ & -2.7755 & 3.054 & 3.093\\
$$& $(0.378 ) $ &$(0.0391 )$ & $(0.00685 )$ & $(3.07  )$  
& $(0.0459 )$  & $(0.000463)$ & $$ & &(0.001)  & ($<0.001$)\\

$4$& $1.020 $ &$1.482 $ & $1.717 $ & $12.43  $  
& $0.3939 $  & $0.01401 $ & $8760.38$ &  -2.8039 & 3.424 & 3.588\\
$$& $(0.378 ) $ &$(0.043 )$ & $(0.00794 )$ & $(1.93  )$  
& $(0.0483 )$  & $(0.00129)$ & $$ &  & ($<0.001$)& ($<0.001$)\\

$5$& $1.13  $ &$1.933 $ & $1.642 $ & $7.495  $  
& $0.5455 $  & $0.004534$ & $8779.88 $ & -2.8102&  3.292 & 3.265\\
$$& $(0.381 ) $ &$(0.0635 )$ & $(0.0100 )$ & $(1.04   )$  
& $(0.086 )$  & $(0.000555 )$ & $$ &  & ($<0.001$) & (0.001)\\

$6$& $1.079  $ &$1.620 $ & $1.590 $ & $5.347  $  & $0.6584 $  & $0.00897$ & $8789.49 $ & -2.8133 & 2.702 & 2.866\\
$$& $(0.379 ) $ &$(0.0481 )$ & $(0.0107 )$ & $(0.542 )$  
& $(0.0726 )$  & $(0.000841 )$ & $$ &  &  (0.003)& (0.002)\\

$7$& $1.006  $ &$1.934 $ & $1.537 $ & $4.431 $  
& $0.9000 $  & $0.003719$ & $8798.81 $ & -2.8162 & 1.809 &  1.727\\
$$& $(0.379 ) $ &$(0.0685 )$ & $(0.00926 )$ & $(0.448 )$  
& $(0.0748 )$  & $(0.000464)$ & $$ &  &  (0.0352)& (0.042)\\

$8$& $1.023 $ &$1.985 $ & $1.500 $ & $3.536  $  
& $0.9173 $  & $0.003220$ & $8802.66$ &  -2.8175
 & 1.058  &  0.940\\
$$& $(0.380 ) $ &$(0.063 )$ & $(0.00933 )$ & $(0.292 )$  
& $( 0.0624 )$  & $(0.000388 )$ & $$ &  & (0.145) & (0.173)\\

$9$& $0.998  $ &$1.882 $ & $1.464 $ & $2.964$  
& $   0.949 $  & $0.003706$ & $8804.44$ & -2.8181& & \\
$$& $( 0.379 ) $ &$(0.112 )$ & $(0.00927 )$ & $(0.216  )$  
& $(0.0484 )$  & $(0.00102 )$ & $$ &  &  & \\
\botrule
\end{tabular}}
\tabnote{ML-estimates for the binomial multifractal with estimated standard errors in parantesis. The Vuong-column reports the likelihood ratio statistic with corresponding $p$-value in brackets \cite{Vuong1989}. The null hypothesis is that MSM(9) and MSM(K) have equal Bayesian Information Criteria, with the alternative hypothesis being that the MSM(9) is closer to the true data generating process. The HAC-adjusted version of the Vuong test \cite{Calvet2004b} corrects for heteroschedacity and autocorrelation in the addends.} \label{tab3}
\end{minipage}
\end{center}
\end{table}

\begin{table}
\begin{center}
\begin{minipage}{\linewidth}
\tbl{Level-GARCH}
{\begin{tabular}{ c  c c c  c c c c c}
\toprule	
&$10^3\times\alpha_0$ & $\gamma$  
& $10^5\times a_0$  & $a_1$ & $b$ & $\nu$& $\log L$ & BIC \\
\colrule
\multicolumn{9}{c}{ U.S. Treausury Bill M3}\\
&$0.0904$ & $0.1699 $  
& $0.7092 $  &$ 0.1301  $ & $0.8915 $ & $3.7995 $& $ 22873.03 $ & $-3.2241$\\
&$(0.225) $ & $(0.0496 )$  
& $(0.139)$  & $(0.0092 )$ & $(0.0062 )$ & $(0.134)$& $$ &  \\

\multicolumn{9}{c}{ NIBORM3}\\

&$ 0.9104 $ & $0.9717 $  
& $0.4889$  & $0.20993 $ & $0.80838 $ & $3.3305$& $8783.27$ & -2.8113
 \\
&$(0.397 )$ & $(0.0719 )$  
& $(0.158)$  & $(0.0209 )$ & $(0.0135 )$ & $(0.155 )$& $$ & \\
\botrule
\end{tabular}}
\tabnote{ML estimates for the parameters in the level-GARCH model. Standard deviations are in brackets.}\label{tab4}
\end{minipage}
\end{center}
\end{table}

\begin{table}
\begin{center}
\begin{minipage}{\linewidth}
\tbl{Level-EGARCH}
{\begin{tabular}{ c  c c c c  c c c c c}
\toprule	
&$10^3\times \alpha_0$ & $\gamma$  
& $a_0$  & $a_1$ & $a_2$& $b$ &  $\nu$& $\log L$ & BIC \\
\colrule
\multicolumn{10}{c}{ U.S. Treausury Bill M3}\\
&$0.1039 $ & $0.3441 $  
& $-0.2360 $  & $-0.03127 $ & $0.2322 $& $0.9888 $ &  $3.866 $& $22936.53 $ & -3.2324 \\
&$(0.212 )$ & $(0.0648 )$  
& $(0.0165 )$  & $(0.00634 )$ & $(0.0123 )$& $(0.00154 )$ &  $(0.138 )$&  &  \\
\colrule
\multicolumn{10}{c}{ NIBORM3}\\
&$ 1.120 $ & $0.9101 $  
& $-0.4024 $  & $0.02930 $ & $0.2664 $& $0.9741 $ &  $3.358 $& $8828.90$ & -2.8245 \\

&$(0.396 )$ & $(0.0887 )$  
& $(0.0372 )$  & $(0.0104 )$ & $(0.0188 )$& $(0.00333 )$ &  $(0.156 )$& $$ & 
\\

\botrule
\end{tabular}}
\tabnote{ML estimates for the parameters in the level-EGARCH model. Standard deviations are in brackets.}\label{tab5}
\end{minipage}
\end{center}
\end{table}

%
%

\begin{table}
\begin{center}
\begin{minipage}{70mm}
\tbl{Jump-diffusion}
{\begin{tabular}{c  c c c c c c c c c c }
\toprule
$10^3\times\alpha_0$ & $\gamma$  
& $10^6\times a_0$  & $a_1$ & $b$ & $c$  & $d$ & $\tau$ &$\log L$ & BIC \\
\colrule

\multicolumn{10}{c}{ U.S. Treasury Bill M3}\\


$0.464 $ & $0.1670 $  
& $5.131 $  & $0.0931 $ & $0.8773 $ & $-3.751 $  & $0.2761 $ & $0.08622$ &$ 22597.89 $ & -3.1839 \\
$( 0.232 )$ & $( 0.0374)$  & $(0.867 )$  & $(0.0055)$ & $(0.0062 )$ & $(0.170 )$  & $(0.0271)$ & $( 0.00624 )$ & &  \\
\colrule
\multicolumn{10}{c}{ NIBORM3}\\
$ 1.085 $ & $1.872 $  
& $0.0436 $  & $0.1291 $ & $0.8301 $ & $-1.883 $  & $-0.2119 $ & $0.01127$ &$8579.61 $ & -2.7431
 \\
$(0.417)$ & $(0.0892)$  
& $(0.0198)$  & $(0.0099 )$ & $(0.0111)$ & $(0.328 )$  & $(0.0417 )$ & $(0.00216 )$ & & \\
\botrule
\end{tabular}}
\tabnote{ML estimates for the parameters in the jump-diffusion model. Standard deviations are in brackets.}\label{tab6}
\end{minipage}
\end{center}
\label{symbols}
\end{table}

\begin{table}
\begin{center}
\begin{minipage}{70mm}
\tbl{In-sample model comparison.}
{\begin{tabular}{l  c c c c c}
\toprule
Model & $\dim(\theta) $ & $\log L$& BIC & Vuong & Hac-adj.	 \\	
\colrule 
\multicolumn{6}{c}{ U.S. Treasury Bill M3} \\
MSM(9)& $6 $ & $22955.80$ & -3.2358&  &  \\	
GARCH & $6$ & 22873.03 & -3.2241& -4.064($<0.001$)&  -3.477($<0.001$)	 \\	
EGARCH & $7$ &22936.53& -3.2324& -1.420 (0.078) & -1.361  (0.087) \\	

Jump-diffusion&7&22597.89 &-3.1839 & -9.878\,\,\,\,\,($<0.001$)& -9.240 \,\,\,\,\,($<0.001$)\\

\colrule 

\multicolumn{6}{c}{ NIBORM3}\\
MSM(9) & $6 $ &  $8804.44$  & -2.8181&  & 	 \\
GARCH & $6$ & $8783.27$ &  -2.8113 &  -1.099(0.136)  &  -1.086(0.139) \\	
EGARCH & $7$ & 8828.90  &-2.8245 & 1.231(0.891 )  &1.470 (0.929) \\	

Jump-diffusion&7&8579.61 &-2.7430& -6.964($<0.001$)&-5.765($<0.001$)
\\

\botrule
\end{tabular}}
\tabnote{The Vuong-column reports the test-statistic for differences in BIC. The null hypothesis is that the multifractal and the alternative model are equally good, with the alternative hypothesis being that the multifractal model performs best. The HAC-adjusted column adjusts for heteroschedacity and autocorrelation in the addends. Corresponding $p$-values are in brackets.}\label{tab7}
\end{minipage}
\end{center}
\label{symbols}
\end{table}

\begin{table}
\begin{center}
\begin{minipage}{\linewidth}
\tbl{Linear drift}
{\begin{tabular}{ l  c c c c  c c c c c c}
\toprule	
Model & $10^3\times\alpha_0$& $10^4\times\alpha_1$ \\
\colrule
\multicolumn{3}{c}{ U.S. Treasury Bill M3} \\
MSM(9)&$0.1214\,(0.354)$ & $-0.1905\,(1.04)$ \\	
GARCH &$0.2821\, (0.386 ) $&  $-0.6731\,(1.10) $ \\	
EGARCH &$0.3595\, (0.310)  $ & $-1.3524\,(0.92)$\\	
Jump-diffusion&$0.7074\,(0.397 ) $ & $-0.8695\,(1.15 )  $\\ 
\colrule
\multicolumn{3}{c}{ NIBORM3}\\
MSM(9)&$2.228\,( 0.726 )  $& $-3.201\,(1.60 ) $ \\	
GARCH &$2.424 \,(0.746)  $&  $-3.986\, (1.66) $ \\	
EGARCH &$ 2.515 \,(0.430) $ & $-3.626\,  (0.87)$\\	
Jump-diffusion&$2.621 \,(0.788)$& $-4.053\,( 1.77 )$\\

\botrule
\end{tabular}}
\tabnote{A linear term in the drift term was added and the parameters estimated using ML. In the TBM3 none of the parameters $\alpha_1$ are significantly different from zero. Standard deviations are reported in brackets.} \label{tab8}
\end{minipage}
\end{center}
\end{table}

\newpage ~ \newpage ~ \newpage


  \newcommand{\printfirst}[2]{#1} \newcommand{\singleletter}[1]{#1}
  \newcommand{\switchargs}[2]{#2#1}

\end{document}